\documentclass[sigconf]{acmart}
\usepackage{amsmath}
\usepackage{tikz}
\usepackage{physics}
\usepackage{pgfplots}
\pgfplotsset{compat = 1.15}
\usepackage{quantikz}
\usepackage{caption}
\usepackage{subcaption}
\usepackage{listings}
\usepackage{todonotes}
\usepackage{pgfplotstable}
\usetikzlibrary{calc}
\usepgfplotslibrary{hvlines}

\usepgfplotslibrary{statistics} 

\AtBeginDocument{%
  \providecommand\BibTeX{{%
    \normalfont B\kern-0.5em{\scshape i\kern-0.25em b}\kern-0.8em\TeX}}}

\setcopyright{acmlicensed}
\copyrightyear{2018}
\acmYear{2018}
\acmDOI{XXXXXXX.XXXXXXX}

\acmConference[Conference acronym 'XX]{Make sure to enter the correct
  conference title from your rights confirmation emai}{June 03--05,
  2018}{Woodstock, NY}
\acmISBN{978-1-4503-XXXX-X/18/06}

\begin{document}

\title[QCardEst/QCardCorr: Quantum Cardinality Estimation and Correction]{QCardEst/QCardCorr:\\Quantum Cardinality Estimation and Correction}

\author{Tobias Winker}
\affiliation{%
  \institution{University of Lübeck}
  \city{Lübeck}
  \country{Germany}}
\email{t.winker@uni-luebeck.de}

\author{Jinghua Groppe}
\affiliation{%
	\institution{University of Lübeck}
	\city{Lübeck}
	\country{Germany}}
\email{jinghua.groppe@uni-luebeck.de}

\author{Sven Groppe}
\affiliation{%
	\institution{University of Lübeck}
	\city{Lübeck}
	\country{Germany}}
\email{sven.groppe@uni-luebeck.de}

\renewcommand{\shortauthors}{Winker, et al.}

\newcommand{\vect}[2]{\begin{pmatrix} #1 \\ #2  \end{pmatrix}}

\begin{abstract}

Cardinality estimation is an important part of query optimization in DBMS. We developed a Quantum Cardinality Estimation (QCardEst) approach using Quantum Machine Learning with a Hybrid Quantum-Classical Network. We define a compact encoding for turning SQL queries into a quantum state, which requires only qubits equal to the number of tables in the query. This allows the processing of a complete query with a single variational quantum circuit (VQC) on current hardware. In addition, we compare multiple classical post-processing layers to turn the probability vector output of VQC into a cardinality value.
We introduce Quantum Cardinality Correction QCardCorr, which improves classical cardinality estimators by multiplying the output with a factor generated by a VQC to improve the cardinality estimation. With QCardCorr, we have an improvement over the standard PostgreSQL optimizer of 6.37 times for JOB-light and 8.66 times for STATS. For JOB-light we even outperform MSCN by a factor of 3.47. 

\end{abstract}



\maketitle

\section{Introduction}
Despite the fact that database systems have evolved significantly over the decades, the development of new database optimization techniques remains crucial to address evolving challenges brought by the exponential growth of data volume and complexity. However, the computational efficiency of new algorithms is heavily influenced by hardware capabilities, and classical computers are approaching fundamental computational limitations controlled by Moore's Law and thermal and energy constraints. These limitations are driving the search for new computing paradigms, such as quantum computing, neuromorphic computing, and specialized hardware accelerators utilizing FPGAs~\cite{Werner2013Hardware} and GPUs~\cite{Groth2022HashGPU,Groth2023UpdateHashGPU}.
Researchers from databases and quantum computing communities have begun to explore quantum-based solutions that leverage the unique computational principles of quantum mechanics for database tasks such as transaction scheduling~\cite{bittner2020avoiding, OJCC_2020v7i1n01_Bittner, Groppe2021Grover}, optimization of join orders~\cite{nayak2023constructing, nayak2024quantum,schonberger2023ready,schonberger2022quantum,Winker2023QMLJOO,Franz2024Hype}, index tuning~\cite{Gruenwald2023IndexTuning}, and relational deep learning~\cite{Vogrin2024RDL}. Their approaches utilize either quantum annealers~\cite{Nayak2024QCE} or quantum machine learning~\cite{Winker2023Tutorial} on universal quantum computers.

The studies~\cite{bittner2020avoiding, OJCC_2020v7i1n01_Bittner, Groppe2021Grover} explored quantum-based solutions for transaction scheduling. Bittner et al.~\cite{bittner2020avoiding} and \cite {OJCC_2020v7i1n01_Bittner} formulated transaction scheduling as a Quadratic Unconstrained Binary Optimization (QUBO) problem and used the quantum annealer to search for near-optimal solutions of transaction scheduling by evolving a quantum system toward its lowest-energy state. They showed that quantum annealing can reduce runtime for scheduling transactions under two-phase locking by finding optimal execution orders faster than classical heuristics like simulated annealing. 
Since finding a conflict-free schedule among a vast set of transaction interleavings is computationally intensive,  the work~\cite{Groppe2021Grover} applied Grover’s algorithm \cite{grover1996fast} to search through this space and achieved a quadratic speed-up over classical algorithms.
To reduce the complexity of transaction scheduling, Nitin et al.~\cite{Nayak2025_5} formulated the transaction scheduling problem as iter-QUBO, developed a locking-based sub-problem generation strategy, and adapted the theoretical framework introduced in~\cite{bittner2020avoiding} and \cite {OJCC_2020v7i1n01_Bittner} to the locking mechanism, significantly enhancing the solvability of scheduling instances,

Quantum computing also provides new and promising solutions to join order optimization (JOO). Winker et al.~\cite{Winker2023QMLJOO} apply quantum machine learning to predict the best join order. While their approach retrieves the join order in one step, Franz et al.~\cite{Franz2024Hype} incrementally build the join tree. The study in \cite{schonberger2023ready} suggested a QUBO solution for the JOO in a limited solution space, and the work~\cite{schonberger2022quantum} extended the solution to the full solution space. The authors in \cite{nayak2023constructing} proposed a QUBO formulation to solve join ordering with direct cost estimations for each join, and their evaluation results show that these techniques achieve better performance in finding valid and optimal shots for real-world queries than the work in \cite{schonberger2023ready}. To mitigate the hardware limitations of current QPUs as reported in \cite{schonberger2023ready}, the authors in \cite{nayak2024quantum} extended the work~\cite{nayak2023constructing} by proposing to partition the big search space of QUBO problems into smaller subspaces and thus enabling QPUs to process more complex join optimizations.

Our work aims at exploring the potential of quantum computing to address the cardinality estimation problem. The cardinality estimation problem in join optimization is a critical challenge in database query optimization. It involves predicting the number of rows (cardinality) that a join operation will produce, which directly impacts the choice of join algorithms, join order, and overall query execution plan. Accurate cardinality estimates enable the query optimizer to select efficient plans, while inaccurate estimates can lead to suboptimal performance. In this work, we proposed quantum-based solutions to the cardinality estimation problem. This work contains the following main contributions:  

\begin{itemize}
    \item Creation of a compact query encoding to encode a SQL query joining $n$ tables into $n$ qubits and thus making it useful for current hardware
    \item Introduction of different classical post-processing layers to map the probability vector from measuring a quantum circuit to a value in $\mathbb{R}$ to use in regression. Comparison of the range of values and probability distributions of these classical layers.
    \item Proposal of cardinality correction to improve the result of an existing classical cardinality estimator using a quantum computer
\end{itemize}

\section{Related Work}



We provide in Table~\ref{tab:RWCardEst} an overview of existing approaches  (classical and quantum) to cardinality estimation, which we discuss in the following subsections in more detail.

\begin{table}
\caption{Existing approaches to cardinality estimation}\label{tab:RWCardEst}
\begin{tabular}{ l c c c c c }
 \textbf{Approach} & \textbf{Trad.} &
 \multicolumn{2}{c}{\textbf{Driven by}} &  \textbf{Quantum}&\textbf{Corr.}\\ 
 &&\textbf{Query} & \textbf{Data}&\\\hline\hline
 PostgreSQL \cite{PostgreSQLGO} & \checkmark &&& \\ 
 MultiHist \cite{10.5555/645923.673638} & \checkmark &&& \\\hline
 MSCN \cite{MSCN} && \checkmark && \\  
 LW-XGB \cite{LW-XGB} && \checkmark && \\\hline
 Naru \cite{Naru} &&& \checkmark & \\  
 DeepDB \cite{DeepDB} &&& \checkmark & \\  
 Flat \cite{FLAT} &&& \checkmark & \\\hline
  UAE \cite{UAE} && \checkmark & \checkmark & \\\hline
  SQL2Circuits \cite{uotila2023sql2circuits} && \checkmark && \checkmark \\
  QardEst \cite{Kittelmann2024Card} && \checkmark && \checkmark \\\hline
  \multicolumn{6}{l}{\textbf{Our approaches:}}\\
  \textbf{QCardEst} && \checkmark && \checkmark \\
  \textbf{QCardCorr} && \checkmark && \checkmark & \checkmark \\
\end{tabular}
\end{table}

\subsection{Classical Cardinality Estimation}

Many cardinality estimation approaches exist, but they can be mainly divided into 4 categories.

\paragraph{Traditional:} Traditional cardinality estimation approaches (e.g., \cite{PostgreSQLGO,10.1145/335168.335230,10.1145/1007568.1007602,10.5555/645923.673638}) rely on assumptions that simplify the properties of real-world queries and data, like column independence. These assumptions often cause substantial estimation errors.

\paragraph{Query driven ML:}

Approaches like MSCN \cite{MSCN} and LW-XGB \cite{LW-XGB} use a set of queries and their true cardinalities to train a machine learning model. These models represent the mapping of query to expected cardinality without necessarily modeling the underlying data.

\paragraph{Data driven ML:}

Approaches like Naru \cite{Naru}, DeepDB \cite{DeepDB}, and Flat \cite{FLAT} use data samples to learn the distribution over the attributes in the database. From these distributions we can calculate the expected cardinality of a query. In comparison to query driven methods, we don't need a set of queries for learning, but the resulting model will be much larger, because we have to model the complete underlying database.

\paragraph{Hybrid:} Hybrid approaches like UAE \cite{UAE} combine query-driven with data-driven machine learning to improve the cardinality estimation further.

\subsection{Quantum Cardinality Estimation}

SQL2Circuits \cite{uotila2023sql2circuits} applies quantum machine learning to cardinality estimation using a quantum natural language processing (QNLP)-inspired approach. This approach encodes queries as parameterized quantum circuit based on category theory using context-free grammar. The parameters of these circuits are then optimized using the Adam optimizer to classify the expected cardinality into one of 4 classes. As this is a classifier, it only outputs in which of a set of predefined intervals the cardinality most likely falls. Our approach in comparison uses regression, which allows it to output an exact expected value for the cardninality estimate.  

QardEst \cite{Kittelmann2024Card} creates a quantum variant of MSCN by replacing classical networks with small quantum circuits. They use the same query encoding as MSCN, which is passed to many individual quantum circuits, which are combined with average pooling. In contrast, we want to use a single quantum circuit for cardinality estimation. This requires a more compact query encoding than MSCN to be feasible on current hardware.

Furthermore, none of the existing approaches proposes to correct an estimated cardinality by a subsequent step, like we propose with the QCardError approach.

\section{Background}



\subsection{Quantum Computing}

Quantum computers are computers that use the effects of quantum mechanics to solve problems. In quantum mechanics, there exist multiple properties that do not occur in classical mechanics. One property is superposition, which allows a quantum system to be not just in one state, but a combination of multiple states. As this is not possible for classical computers, the state of the quantum system collapses into one of the states of the superposition on measurement. Another property is entanglement, which makes the states of two qubits dependent on each other. The goal of quantum computing is to exploit these phenomena to solve problems more efficiently than a classical computer could. 

The state of a quantum computer can be described by using vectors of complex numbers.
The minimal unit of a quantum computer is the qubit. A qubit has two basis states $\ket{0}$ and $\ket{1}$ corresponding to the two states of a classical bit. Contrary to a classical bit, a qubit can also be in a superposition of these basis states. A superposition is a linear combination of the two basis states $\ket{0}$ and $\ket{1}$: 
$$
\psi = a_0\cdot\ket{0}+a_1\cdot\ket{1} 
$$
where $a_0$ and $a_1$ are complex numbers and satisfy the condition $|a_0|^2+|a_1|^2=1$. The values $a_0$ and $a_1$ are called amplitudes, and when a qubit is measured, it collapses into the state $\ket{0}$ with probability $|a_0|^2$ and into the state $\ket{1}$ with probability $|a_1|^2$.
A quantum state consisting of multiple qubits can be described in the same way. For $n$ qubits, the quantum state has $2^n$ basis states and is defined by $2^n$ amplitudes as

$$
\psi = \sum_{i=0}^{2^n-1} a_i \cdot \ket{i}
$$
with $\sum_{i=0}^{n^2-1} |a_i|^2=1$. As a quantum state is completely defined by its amplitudes, it can be written as a vector. The basis states of a single qubit in vector form are $\ket{0}=\vect{1}{0}$ and $\ket{1}=\vect{0}{1}$ and general state of a single qubit is the superposition $a_0\cdot\ket{0}+a_1\cdot\ket{1}=\vect{a_0}{a_1}$.   

\subsection{Quantum Gates}

The quantum state is changed by applying a unitary operator to it. As the states can be written as vectors, the operators can be written as unitary matrices. They have to be unitary because they have to be reversible, and the length of the vector cannot change. For example, the NOT operation, which turns $\ket{0}$ into $\ket{1}$ and vice versa, has the matrix form 
$X=
\begin{pmatrix}
  0 & 1\\ 
  1 & 0
\end{pmatrix}
$
as can be seen by:
$$
X\cdot \ket{0} = \begin{pmatrix} 0 & 1\\ 1 & 0\end{pmatrix} \vect{1}{0}
=\vect{0}{1}=\ket{1}
$$

There exists many quantum gates, but we will only introduce the two types that are important for our approach. The first are the rotation gates. These are parameterized gates, which rotate the state of a single qubit around one of the axes of the Bloch sphere.

\newcommand{\mats}[4]{ \begin{bmatrix} #1 & #2 \\ #3 & #4 \end{bmatrix} }
\newcommand{\thetahalf}[0]{ \frac{\theta}{2} }

The second type of gate which are relevant for us is the controlled Pauli gates. A controlled gate acts on multiple qubit with a control qubit and a target qubit. A gate is applied to the target qubit if the control qubit is $\ket{1}$. As the control can be in a superposition, the target qubit will also be in a superposition. As the states of the control and target qubit now depend on each other they are entangled. An example for such a gate is the controlled NOT (CNOT gate), which applies the NOT gate to the second qubit, if the first qubit is in the state $\ket{1}$..

\subsection{Hybrid Quantum Computing}

As the number of qubits of current QPUs is still limited, these devices are often used for hybrid algorithms. A hybrid quantum-classical algorithm is an algorithm that runs some parts on a quantum computer and others on a classical computer. As the measurement of a quantum state leads to the collapse of the superposition, a continuous interaction between a classical computer and a quantum circuit is not possible. Instead, a classical algorithm creates a quantum circuit and runs it on a quantum computer like a function call.

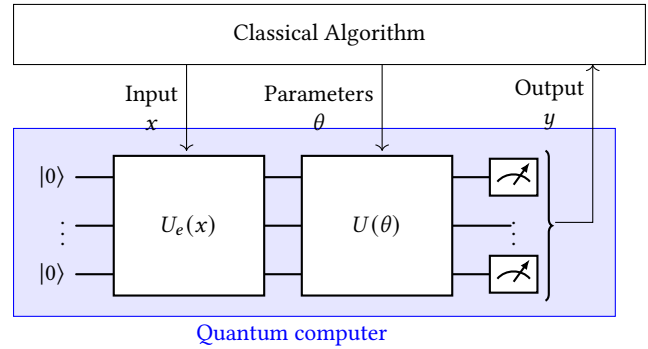
\begin{figure}
    \centering
    \begin{tikzpicture}
    \node[rectangle,
    minimum height=2.5cm,
    minimum width=8cm,
    fill= blue!10,
    text= blue,
    draw = blue] (r) at (0.3,0.5) {};
    \node[text=blue] at (0,-1) {Quantum computer};
    \node at (0,0.5) (qc) {\begin{tikzcd} 
            [row sep=0.1cm]
            \lstick{$\ket{0}$} & \gate[3, nwires=2][2cm]{U_e(x)} & \gate[3, nwires=2][2cm]{U(\theta)} & \meter{}  \rstick[wires=3]{} \\
            \lstick{\vdots} &  & &  \vdots \\
            \lstick{$\ket{0}$} &  & &  \meter{} 
    \end{tikzcd}};
    \node at (0.5,3) [rectangle,draw,
            minimum height = 0.8cm,
            minimum width = 8.4cm] (classical) {Classical Algorithm};
    \draw [->] (-1.4,2.6) -- (-1.4,1.45) node [align=center,left,pos=0.5] {Input \\ $x$}; 
    \draw [->] (1.2,2.6) -- (1.2,1.45) node [align=center,left,pos=0.5] {Parameters \\ $\theta$}; 
    \draw [->] (3.5,0.5) -- (4,0.5) -- (4,2.6) node [align=center,left,pos=0.75] {Output \\ $y$};
\end{tikzpicture}
    \caption{Quantum Classical Hybrid Approach using a VQC}
    \label{fig:hybrid}
\end{figure}

In quantum machine learning, we use a hybrid approach by using a quantum circuit as the machine learning model and using a classical computer to optimize the parameters of the quantum circuit, as seen in Fig. \ref{fig:hybrid}. A quantum circuit that is optimized this way is called a variational quantum circuit (VQC).

\subsection{Variational Quantum Circuits}

A VQC requires three parts: an encoding layer, a computation layer, and a layer containing the measurement of the qubits. The encoding layer turns the classical input data into a quantum state, which represents the data. It is described by an operator $U_e(x)$, which is a parameterized circuit depending on the input parameters $x$. encoding methods exist \cite{10.1049/qtc2.12032}, which differ in the type of data they can encode, the required number of qubits, and the depth of the encoding circuit. We will use rotation encoding as it allows the encoding of two values into a single qubit and requires only a single rotation gate per value. 

The computation layer $U(\theta)$ turns the quantum state representing the input data, created by the encoding layer, into a quantum state representing the output data based on the parameters $\theta$. Finally, the measurement is used to collapse the superposition of the output state into a classical bit string. By running the circuit multiple times we get a probability vector for the different quantum states as output.

\section{Approach}

In this section, we will introduce our quantum encoding of a query, cardinality correction, and different classical post-processing layers, which we will compare.

\subsection{Encoding Queries}

To use quantum computing for processing queries, we have to define an encoding that creates a quantum state based on the query. We have multiple qualities of the encoding we have to consider:

\paragraph{Uniqueness:}
Each possible query should be encoded into a different quantum state. If two different queries were encoded into the same state, some information about the query was lost in encoding. The model would produce the same result for both queries, even through their cardinalities can be different. In a mathematical sense, we want the encoding operation $U_e(x)$ to be an injective function.

\paragraph{Qubit complexity:}
As the number of qubits on current QPUs and classical simulators is limited, our encoding should use as few qubits as possible.

\paragraph{Encoding circuit depth:}
A deeper encoding circuit will lead to a deeper circuit overall, which causes more noise. This makes a deep encoding less useful for current QPUs. 
\\
\\
We work with SQL queries of the form

\begin{lstlisting}[language=SQL]
    SELECT * FROM table1, table2, table3
    WHERE table1.pKey=table2.key1
    AND table1.pKey=table3.key1
    AND table2.colA > 100
    AND table3.colB = 10
\end{lstlisting}

The information in the query consists of 
\begin{itemize}
    \item \emph{Columns:} A list of columns that should be returned or * that represents all columns.
    \item \emph{Tables:} A list of tables that will be used in this query.
    \item \emph{Join conditions:} A list of conditions in the form \\ tableA.column=tableB.column. These conditions will be used in a join to correlate the entries from two tables,
    \item \emph{Filter conditions:} A list of conditions of the form table.column <operator> <constant>. These remove entries that do not fulfill the condition from the result, and thus reducing the cardinality of the result. 
\end{itemize}

As the cardinality of a query is the number of returned rows, the selected columns do not affect the cardinality, and this part of the query can be ignored.
The tables can be mapped to the ids $t_i$.
By limiting ourself to primary-foreign key joins, which are the most common joins used in real-world database applications, we can ignore the join conditions.

The filter conditions have a significant effect on the cardinality of the query as they limit the number of rows in the table, that are used. As the main factor for the cardinality is the size of the joined tables, we can combine all filters affecting a table $t_i$ into a single selectivity value $s_i\in [0,1]$, which describes the percent of rows, which pass all filters applied to a table.

Thus, we reduced a query of $n$ tables to $2n$ values $t_i$ and $s_i$. We can encode these values uniquely into $n$ qubits, by using rotation encoding for $t_i$ around one axis and a rotation around a different axis for $s_i$. As the maximal values for both $t_i$ and $s_i$ are known, the number of tables in the database for $t_i$ and 1 for $s_i$, we can scale them accordingly to keep the angle of rotation lower than the full rotation around the Bloch sphere.

\begin{figure*}[ht!]
	\captionsetup[subfigure]{position=b}
	\begin{subfigure}{0.48\columnwidth}
		\includegraphics[width=\columnwidth]{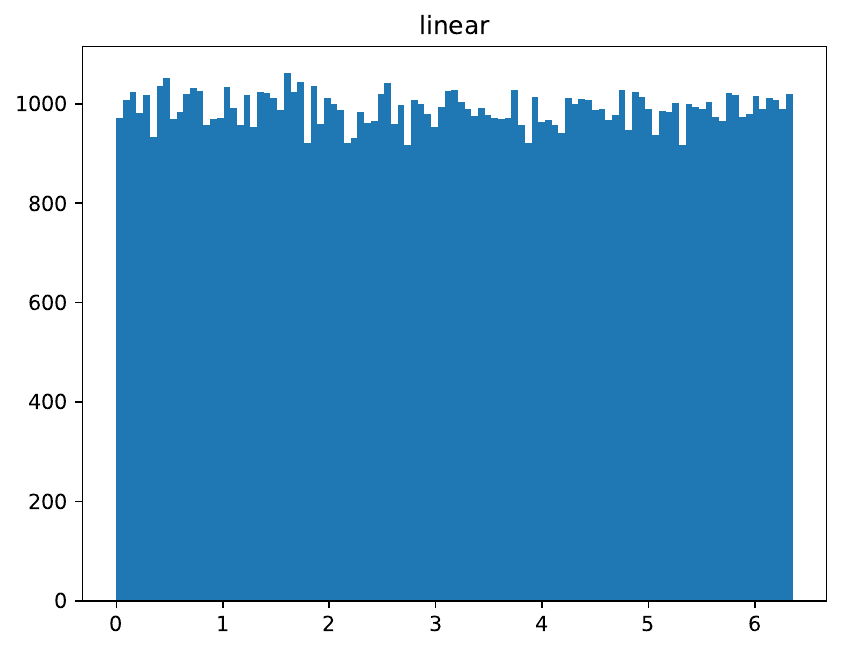}
		\caption{Linear layer}
	\end{subfigure}
	\begin{subfigure}{0.48\columnwidth}
		\includegraphics[width=\columnwidth]{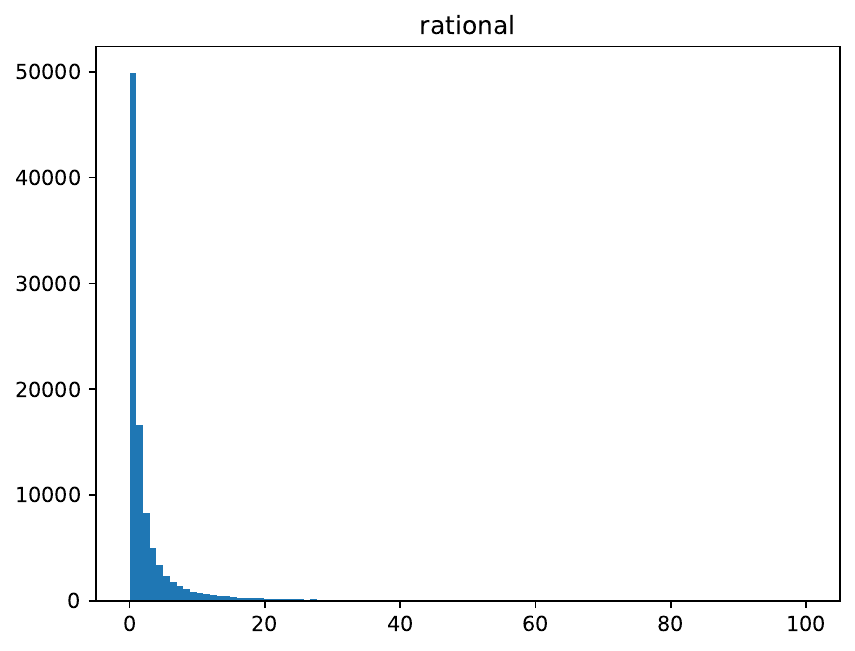}
		\caption{Rational layer}
	\end{subfigure}
	\begin{subfigure}{0.48\columnwidth}
		\includegraphics[width=\columnwidth]{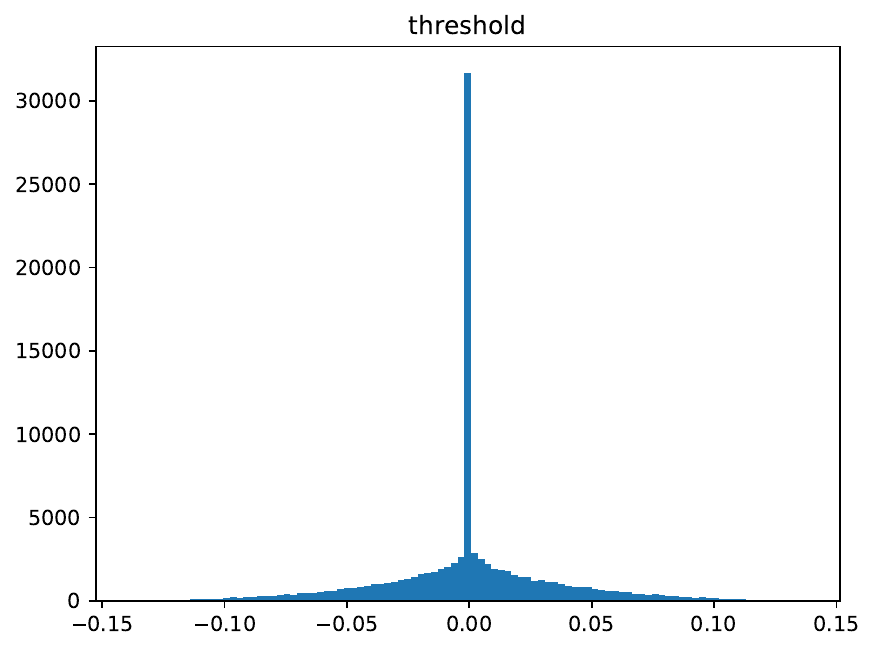}
		\caption{Threshold layer}
	\end{subfigure}
	\begin{subfigure}{0.48\columnwidth}
		\includegraphics[width=\columnwidth]{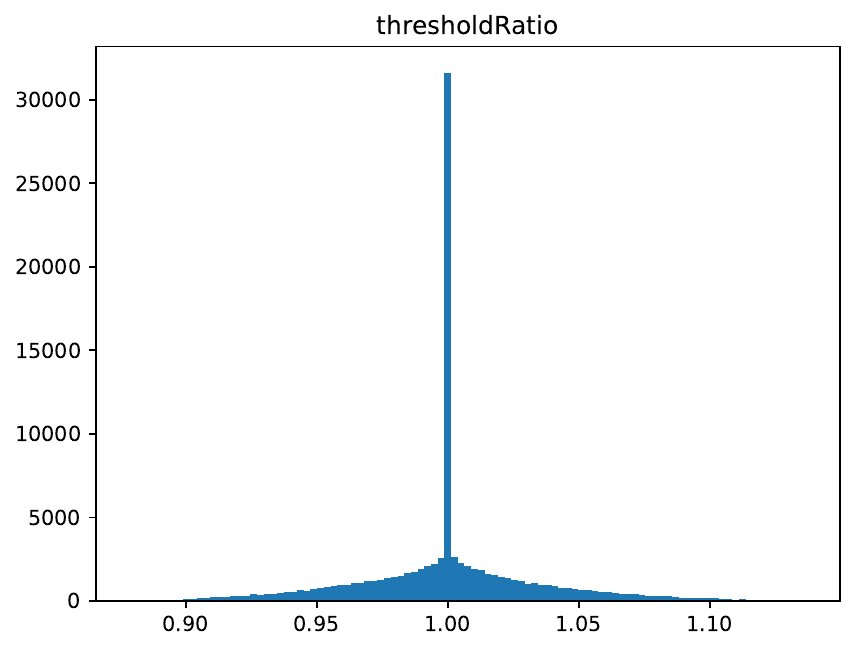}
		\caption{Rational threshold layer}
	\end{subfigure}
	\begin{subfigure}{0.48\columnwidth}
		\includegraphics[width=\columnwidth]{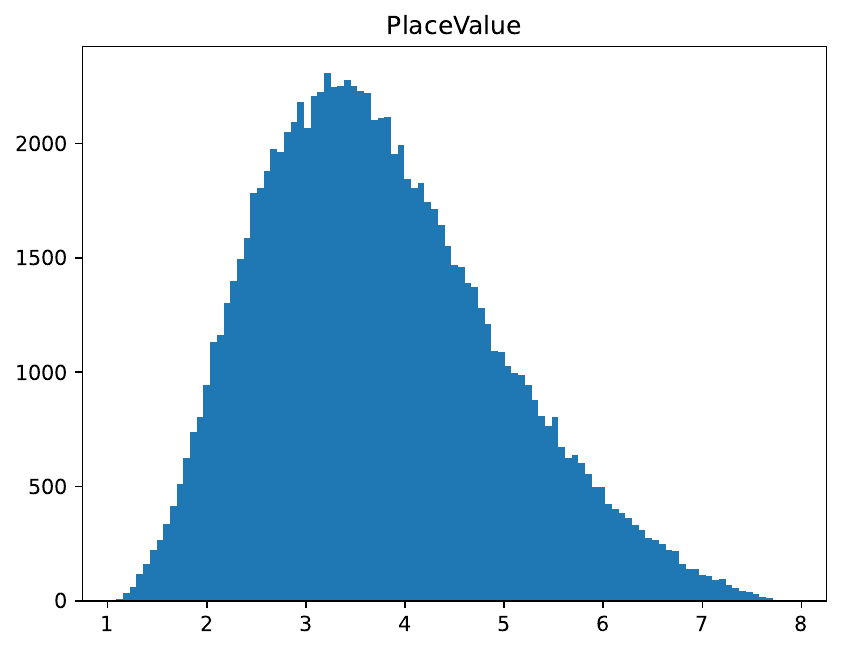}
		\caption{Place Value layer with 4 qubits \newline}
	\end{subfigure}
	\begin{subfigure}{0.48\columnwidth}
		\includegraphics[width=\columnwidth]{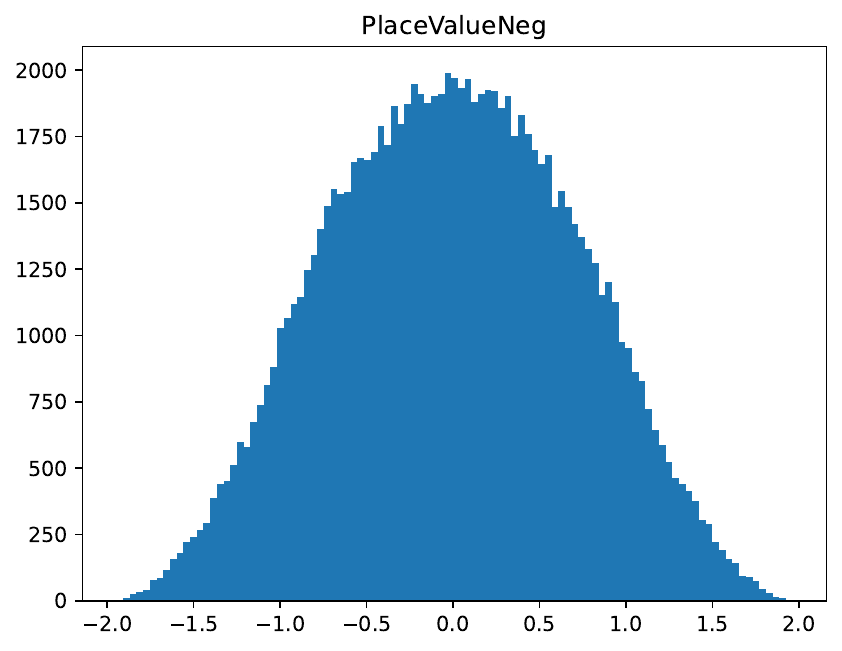}
		\caption{Place Value layer with 4 qubits and with negative Values}
	\end{subfigure}
	\begin{subfigure}{0.48\columnwidth}
		
		\includegraphics[width=\columnwidth]{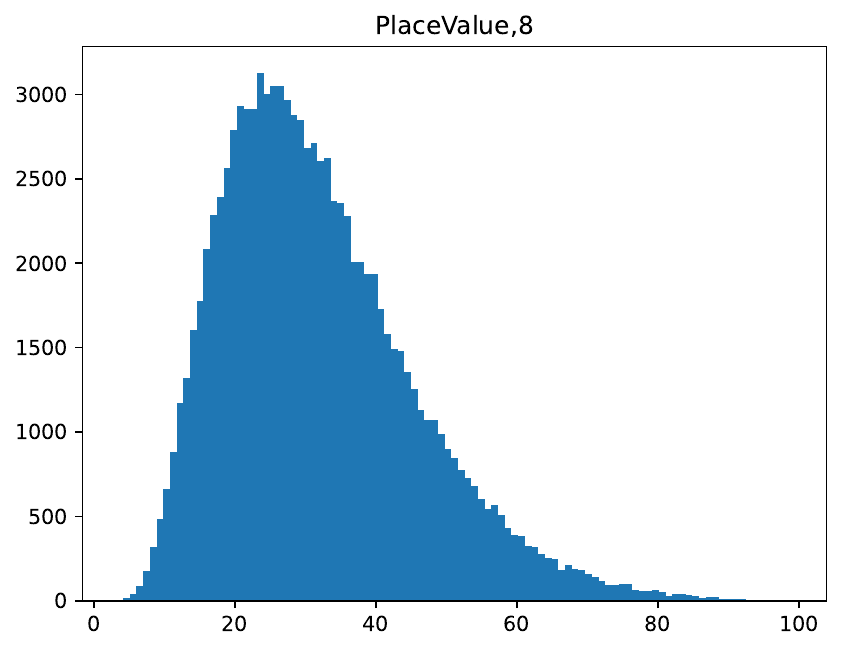}
		\caption{Place Value layer with 8 qubits}
	\end{subfigure}
	\begin{subfigure}{0.48\columnwidth}
		\includegraphics[width=\columnwidth]{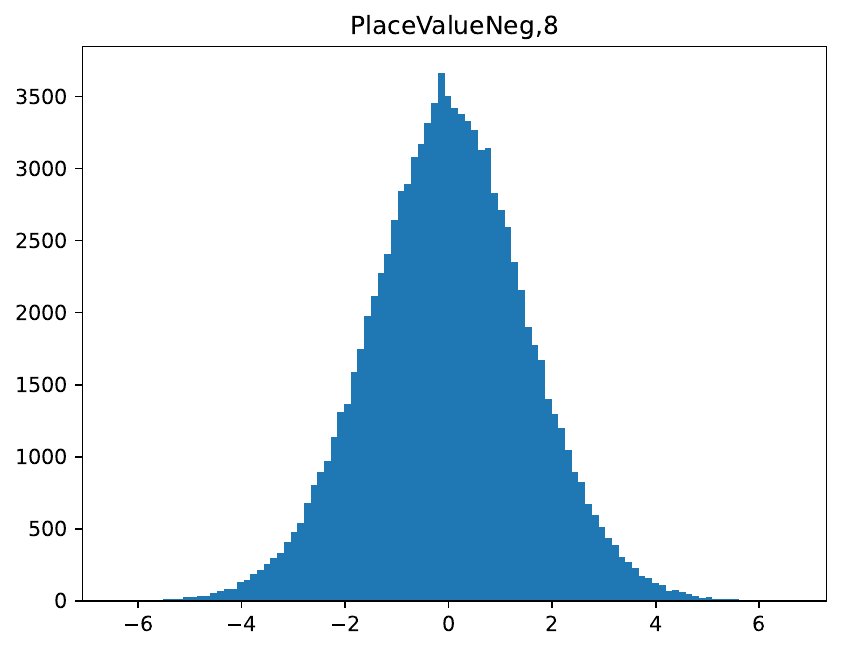}
		\caption{Polynomial layer with negative Values}
	\end{subfigure}
	\caption{Value distribution of different classical post-processing layers. Scalars are set to the constant 1, and PlaceValue uses the base 2}
	\label{fig:valueDistribution}
\end{figure*}

\subsection{Quantum Cardinality Correction}

We use a query-driven approach to use quantum machine learning for cardinality estimation. Another alternative is quantum cardinality correction, where we predict the cardinality with a classical approach and then use quantum machine learning to create a factor to improve the quality of the cardinality estimation.

Let $q$ be a query and $f_p(q)$ be a cardinality estimation model with parameters p, $t(q)$ the true cardinality of the query $q$, and an error function $e$. In cardinality estimation, our goal is to minimize the error over all queries in our training set $Q$:

$$
\min_{p} \sum_{q\in Q} e(f_p(q),t(q))
$$

For cardinality correction, we use a second model $g_x(q)$ which corrects the prediction of the model $f$ for fixed parameters.

$$
\min_{x} \sum_{q\in Q} e(f(q)*g_x(q),t(q))
$$

The model $g$ would be optimal, if $g(q)=\frac{t(q)}{f(q)}$ as 
$f(q)*g(q) = f(q) * \frac{t(q)}{f(q)} = t(q)$. Thus, the optimization goal is:

$$
\min_{x} \sum_{q\in Q} e\left(\frac{t(q)}{f(q)},g(q)\right)
$$

If the classical cardinality estimation model is good, the cardinality correction model has to output a value close to 1 for most queries. While the outputs for cardinality estimation are natural numbers $\mathbb{N}^+$, the outputs of cardinality correction are positive rational numbers $\mathbb{Q}^+$. 
If the classical model over- and underpredicts with similar probability, the amount of correction values in the interval $[0,1]$ will be similar to the interval $[1,\infty]$. By applying a logarithm to the factors of cardinality correction, the value range is converted to the real numbers $\mathbb{R}$, and 0 will be the neutral value of correct predictions.

\subsection{Classical Layer}



We use a hybrid model, which consists of a VQC followed by a classical layer. The task of this classical layer is the conversion of the output vector of the VQC into a single value. This is necessary to achieve the desired output values as the output of the VQC is limited to a vector of probabilities $x$ and thus all individual values in the vector are limited to $x_i\in[0,1]$ with the additional condition of $\sum x_i = 1$.
Here we will now introduce a number of different functions for this classical layer. These will get the output vector $x$ from the VQC as input and produce a single value $v$. The number of required outputs from the VQC varies for the different functions. Additional, this classical layer could have its own parameters $s$ to scale the output values, which will be optimized together with the parameters of the VQC. As we want to use a gradient descent optimizer, the function of the classical layer have to be derivable.

\paragraph{Linear}

This is the simplest way of turning the probability vector into a single value. We just take the first value of the vector and multiply it by a scalar.
\begin{equation}\label{eq:linear}
    v = x_0*s
\end{equation}
We can use the maximum value of $y_{max}$ of the desired outputs $y$ as the scaling factor, resulting in a value range of $v\in[0,y_{max}]$. This requires that the maximal possible output value is known. Alternatively, $s$ can be a parameter which is optimized together with the parameters of the VQC.

\paragraph{Rational}

In this layer we use the ratio of two probabilities as our function. To assure that the function is continuous, we have to add a small value $\epsilon$ to the probabilities, preventing a division by zero.
\begin{equation}\label{eq:rational}
    v = \frac{x_0+\epsilon}{x_1+\epsilon}
\end{equation}
The value range of this is $v\in \mathbb Q^{0}$. Thus, in theory, this layer can produce any positive rational number, but the numbers close to 1 are more likely.

By applying a logarithm to this value we can turn the range into $w\in \mathbb R$. Here the values close to 0 are more likely.
\begin{equation}\label{eq:rationalLog}
    v =  \log\frac{x_0+\epsilon}{x_1+\epsilon}
\end{equation}

\paragraph{Threshold}

For the previously introduced layers, it was always hard to hit a specific value. We now propose a special layer that has an exact predefined value for many cases. For this, we use the $relu(x)$ function, which is often used in neural networks as an activation function \cite{relu}. This function is 0 for all negative inputs and linear for positive inputs. By adding an offset $d$, we have a function that is 0 for $x<=d$. In this way, we can use one probability value $x_i$ to control if another value $x_j$ affects the value $v$ by combining them as 
$$v = relu(x_i-d)*x_j*s^2$$ 
with a constant $d\in[0,1]$ as the threshold and a learned scalar s. The scalar is squared to have a larger input as the maximal value without it is limited to $relu(x_i-d)*x_j<=(\frac{1-d}{2})^2<=\frac{1}{4}$

To allow a higher range, we combine two of these threshold-controlled values by either subtraction or division.
\begin{equation}\label{eq:thresholdDiff}
    v = relu(x_0-d)*x_1*s_1^2 - relu(x_2-d)*x_3*s_2^2
\end{equation}
\begin{equation}\label{eq:thresholdFactor}
    v =\frac{1+relu(x_0-d)*x_1*s_1^2}{1+relu(x_2-d)*x_3*s_1^2}
\end{equation}

\paragraph{Place Value}

This layer is inspired by place-value notation, such as the binary or decimal system. Like in these systems, every position in the string of digits has an increasingly higher value. In opposition to the traditional place-value system the digits are not from a limited set of options (e.g., from $\{0,1\}$ for the binary system), but are a continues value in the interval $[0,1]$. In addition to the inputs $x$, we also require a base $b>1$. Similarly to the previous layers, this scalar can be fixed or be optimized together with parameters of the VQC. This layer can work with a variable number of qubits. Let $l$ be the number of qubits, then $v$ is calculated as:
\begin{equation}\label{eq:PlaceValue}
    v = \sum_{i=0}^{l-1} x_i*b^i
\end{equation}
This can be extended to allow for negative numbers by using half of the qubits with a negative factor.
\begin{equation}\label{eq:PlaceValueNeg}
    v = \sum_{i=0}^{\frac{l}{2}-1} x_i*b^i - \sum_{i=0}^{\frac{l}{2}-1} x_{\frac{l}{2}+i}*b^i
\end{equation}

\subsection{Value Distribution}

We examine the distribution of values created by the different functions. For this, we create random quantum states based on the Haar measure. The measurement probabilities of these random quantum states are taken as simulated outputs of the VQC and used as input for the classical layer. Figure \ref{fig:valueDistribution} shows the resulting value distributions as histograms. For the two threshold layers, we can clearly see how most quantum states will result in the neutral value. The rational layer has a similar spike, but in this case, it is not always the same value, and instead, many values close together land in the same histogram bucket. For the PlaceValue layers, we can see that the distribution gets thinner with more qubits, but gains an overall larger value range.

\section{Evaluation}

\subsection{Setup}

We used a VQC with 6 qubits and 16 layers. Each layer consists of a circular encoding layer using the CY gate, followed by a parameterized RY and RZ gate.  The table ids are encoded by a rotation around the x-axis, and the selectivities are encoded by a rotation around the z-axis. The optimization is run for 8000 episodes using the Adam optimizer with a decaying learning rate. For the cardinality correction, the classical cardinalities to be corrected are taken from the standard PostgreSQL optimizer. 
The approach is implemented in Python (v. 3.10.6) and qiskit (v. 0.39.2) for the simulation of the quantum circuit and optimized with PyTorch (v. 2.2.2). 
We used both the JOB-light \cite{10.14778/2850583.2850594} benchmark for the IMDB database with 70 queries as well as the STATS \cite{han2021cardinality} benchmark with 142 queries, with up to 6 tables joined in a single query.

We used a total of 9 different classical post-processing layers:

\begin{itemize}
    \item \emph{Linear} from equation \ref{eq:linear} with the maximal cardinality in the training data as $s$
    \item \emph{Rational} from equation \ref{eq:rational} 
    \item \emph{RationalLog} from equation \ref{eq:rationalLog} 
    \item \emph{Threshold} from equation \ref{eq:thresholdDiff} 
    \item \emph{ThresholdRatio} from equation \ref{eq:thresholdFactor} 
    \item \emph{PlaceValue} from equation \ref{eq:PlaceValue} with $l=4$
    \item \emph{PlaceValue8} from equation \ref{eq:PlaceValue} with $l=8$
    \item \emph{PlaceValueNeg} from equation \ref{eq:PlaceValueNeg} with $l=4$
    \item \emph{PlaceValueNeg8} from equation \ref{eq:PlaceValueNeg} with $l=8$ 
\end{itemize}

Each layer is used once for cardinality estimation and once for cardinality correction for each of the two datasets.

\definecolor{darkgreen}{rgb}{0.0, 0.5, 0.0}
\begin{figure}[t]
	\includegraphics[width=\columnwidth]{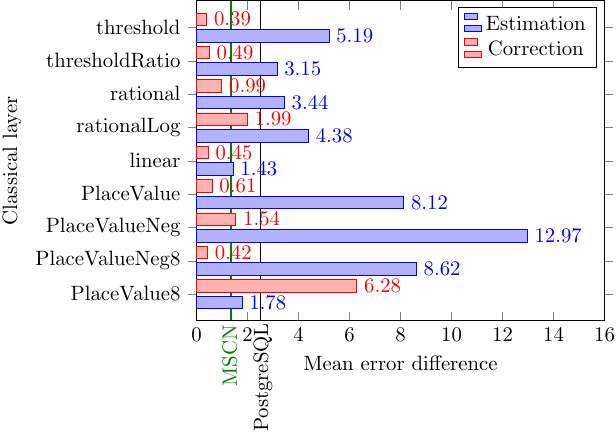}
	\caption{JOB-light benchmark: The black line is the cardinality estimation of PostgreSQL and the green line the cardinality estimation of MSCN}
	\label{fig:jobResults}
\end{figure}
\begin{figure}[t]
	\includegraphics[width=\columnwidth]{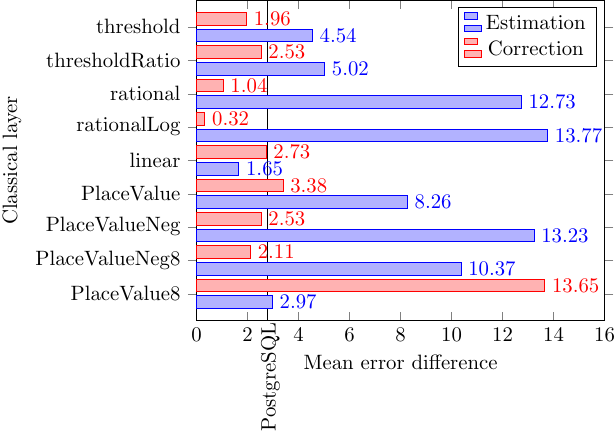}
	\caption{STATS benchmark: The black line is the cardinality estimation of PostgreSQL}
	\label{fig:statsResults}
\end{figure}

\subsection{Results}

Fig.~\ref{fig:jobResults} shows the results for the JOB-light benchmark and Fig.~\ref{fig:statsResults} for the STATS benchmark. As an evaluation metric, we use the average difference between predicted cardinality and the corrected PostgreSQL cardinality. The cardinalities in this case are the logarithmic values of the cardinalities, because otherwise the estimation errors for queries with large cardinalities will have a major effect on the average.

For the JOB-light benchmark, the overall best approach is cardinality correction using the Threshold layer, being 6.37 times better than the PostgreSQL estimator and 3.47 times better than MSCN\footnote{Please note that according to \cite{Kittelmann2024Card}, the MSCN approach with
256 neurons in each hidden layer (like we use in our evaluation) outperforms all assessed quantum neural networks in the QardEst approach.}. The second best is the PlaceValue8Neg layer, being 5.91 times better than PostgreSQL and 3.22 times better then MSCN. In 6 of 8 cases where cardinality correction is better than PostgreSQL it is also better than MSCN.

For the STATS benchmark, on the other hand, the Rational and RationalLog layers are the best, with 8.66 and 2.67 times better than PostgreSQL.

For the direct cardinality estimation, only Linear and PlaceValue8 perform better, than PostgreSQL. For the cardinality correction, only PlaceValue8 is much worse than PostgreSQL. This stems from the fact that these can only produce positive numbers, and thus can only correct the cardinality by increasing it. While PlaceValue8 is the worst for cardinality correction, it is one of the best for cardinality estimation, and its variant PlaceValueNeg8, which allows for negative numbers, is one of the best for cardinality correction.

\section{Conclusions}

We have presented a compact quantum encoding for SQL queries and classical post-processing layers. We have shown that with these methods a quantum cardinality estimator can be better than the PostgreSQL cardinality estimation. Finally, we used a quantum circuit for cardinality correction to significantly improve the cardinality estimations. This shows great potential for hybrid quantum-classical algorithms. 

\section*{Artifacts}

The code can be found in the GitHub repository:\\
https://github.com/TobiasWinker/QCardEst. \\
The README file in the repository describes the replication of the results.

\bibliographystyle{ACM-Reference-Format}

\end{document}